\documentclass[runningheads]{llncs}
\usepackage{graphicx}

\usepackage{amsfonts}

\usepackage{xcolor} 

\usepackage{epstopdf}

\usepackage{amsmath}

\usepackage{xcolor}

\usepackage[shortlabels]{enumitem}

\usepackage{caption}

\usepackage{array}
\newcolumntype{L}[1]{>{\raggedright\arraybackslash}p{#1}}
\newcolumntype{C}[1]{>{\centering\arraybackslash}p{#1}}
\newcolumntype{R}[1]{>{\raggedleft\arraybackslash}p{#1}}

\usepackage[backend=biber,
sorting=nyt,
url=false,
doi=false,
maxnames=4,
minnames=1,
style=lncs]{biblatex}

\DeclareNameFormat{author}{%
  \ifdefvoid{\namepartprefix}{}{\namepartprefix\space}\namepartfamily, \namepartgiveni%
  \ifthenelse{\value{listcount}<\value{liststop}}
    {\addcomma\space}%
    {}%
  \usebibmacro{name:andothers}%
}
\DeclareNameFormat{editor}{%
  \ifdefvoid{\namepartprefix}{}{\namepartprefix\space}\namepartfamily, \namepartgiveni%
  \ifthenelse{\value{listcount}<\value{liststop}}
    {\addcomma\space}%
    {\space\ifthenelse{\value{listcount}>1}
      {(\bibstring{editors})}
      {(\bibstring{editor})}}%
}
\DeclareBibliographyAlias{misc}{article}
\DeclareFieldFormat[article]{volume}{\textbf{#1}}
\usepackage{pifont}
\newcommand{\cmark}{\ding{51}} 
\newcommand{\xmark}{\ding{55}} 

\definecolor{yelloworange}{HTML}{FFC000}

\newcommand\figref{Fig.~\ref}
\newcommand\tabref{Table~\ref}
\newcommand\secref{Section~\ref}

\usepackage{hyperref}

\begin{document}

  
%
\title{Adaptive Multi-scale Online Likelihood Network for AI-assisted Interactive Segmentation}
%
\titlerunning{Adaptive Multi-scale Online Likelihood Network}
\ifdefined\ANONYMIZED
\author{Anonymous MICCAI Submission, Paper 2125}
\else
\author{Muhammad Asad\inst{1}
\and
Helena Williams \inst{2}
\and
Indrajeet Mandal\inst{3} \and
Sarim Ather \inst{3} \and
Jan Deprest \inst{2} \and
Jan D'hooge \inst{4} \and
Tom Vercauteren \inst{1}
}
\authorrunning{M Asad et al.}

%
\institute{School of Biomedical Engineering \& Imaging Sciences, King’s College London, UK \and Department of Development \& Regeneration, KU Leuven, Belgium \and Radiology Department, Oxford University Hospitals NHS Foundation Trust, UK \and
Department of Cardiovascular Sciences, KU Leuven, Belgium
}
\fi

\maketitle              
\begin{abstract}
Existing interactive segmentation methods leverage automatic segmentation and user interactions for label refinement, significantly reducing the annotation workload compared to manual annotation. However, these methods lack quick adaptability to ambiguous and noisy data, which is a challenge in CT volumes containing lung lesions from COVID-19 patients. In this work, we propose an adaptive multi-scale online likelihood network (MONet) that adaptively learns in a data-efficient online setting from both an initial automatic segmentation and user interactions providing corrections. We achieve adaptive learning by proposing an adaptive loss that extends the influence of user-provided interaction to neighboring regions with similar features. In addition, we propose a data-efficient probability-guided pruning method that discards uncertain and redundant labels in the initial segmentation to enable efficient online training and inference. Our proposed method was evaluated by an expert in a blinded comparative study on COVID-19 lung lesion annotation task in CT. Our approach achieved 5.86\% higher Dice score with 24.67\% less perceived NASA-TLX workload score than the state-of-the-art. Source code is available at: \href{https://github.com/masadcv/MONet-MONAILabel}{https://github.com/masadcv/MONet-MONAILabel}

\end{abstract}
\section{Introduction}
Deep learning methods for automatic lung lesion segmentation from CT volumes have the potential to alleviate the burden on clinicians in assessing lung damage and disease progression in COVID-19 patients \cite{revel2021study,roth2021rapid,rubin2020role}. 
However, these methods require large amounts of manually labeled data to achieve the level of robustness required for their clinical application \cite{chassagnon2020ai,gonzalez2021detecting,wang2020noise,tilborghs2020comparative}.
Manual labeling of CT volumes is time-consuming and may increase the workload of clinicians. 
Additionally, applying deep learning-based segmentation models to data from new unseen sources can result in suboptimal lesion segmentation due to unseen acquisition devices/parameters, variations in patient pathology, or future coronavirus variants resulting in new appearance characteristics or new lesion pathologies \cite{mclaren2020bullseye}. 
To address this challenge,
interactive segmentation methods that can quickly adapt to such changing settings are needed. 
These can be used either by end-users or algorithm developers to quickly expand existing annotated datasets
and enable agile retraining of automatic segmentation models \cite{budd2021survey}.

\subsubsection{Related work:}
Interactive segmentation methods for Artificial Intelligence (AI) assisted annotation have shown promising applications in the existing literature \cite{luo2021mideepseg,rajchl2016deepcut,wang2018deepigeos,wang2018interactive}. 
BIFSeg \cite{wang2018interactive} utilizes a bounding box and scribbles with convolutional neural network (CNN) image-specific fine-tuning to segment potentially \textit{unseen} objects of interest.
MIDeepSeg \cite{luo2021mideepseg} incorporates user-clicks with the input image using exponential geodesic distance.
However, BIFSeg, MIDeepSeg, and similar deep learning-based methods exploit large networks that do not adapt rapidly to new data examples in an online setting due to the elevated computational requirements. 

Due to their quick adaptability and efficiency, a number of existing online likelihood methods have been applied as interactive segmentation methods \cite{asad2022econet,boykovjolly2001interactive,wang2016dynamically}. 
DybaORF \cite{wang2016dynamically} utilizes hand-crafted features with dynamically changing weights based on interactive labels' distribution to train a Random Forest classifier. 
ECONet \cite{asad2022econet} improves online learning with a shallow CNN that jointly learns both features and classifier to outperform previous online likelihood inference methods. 
While ECONet is, to the best of our knowledge, the only online learning method that addresses COVID-19 lung lesion segmentation, it is limited to learning from user scribbles only. This means that it requires a significant amount of user interaction to achieve expert-level accuracy.
Additionally, the model uses a single convolution for feature extraction, limiting its accuracy to a specific scale of pathologies. For each CT volume, the model is trained from scratch, resulting in lack of prior knowledge about lesions.

\ifnum 1=1
\begin{figure}[t!]
  \centering
  \includegraphics[width=0.8\linewidth]{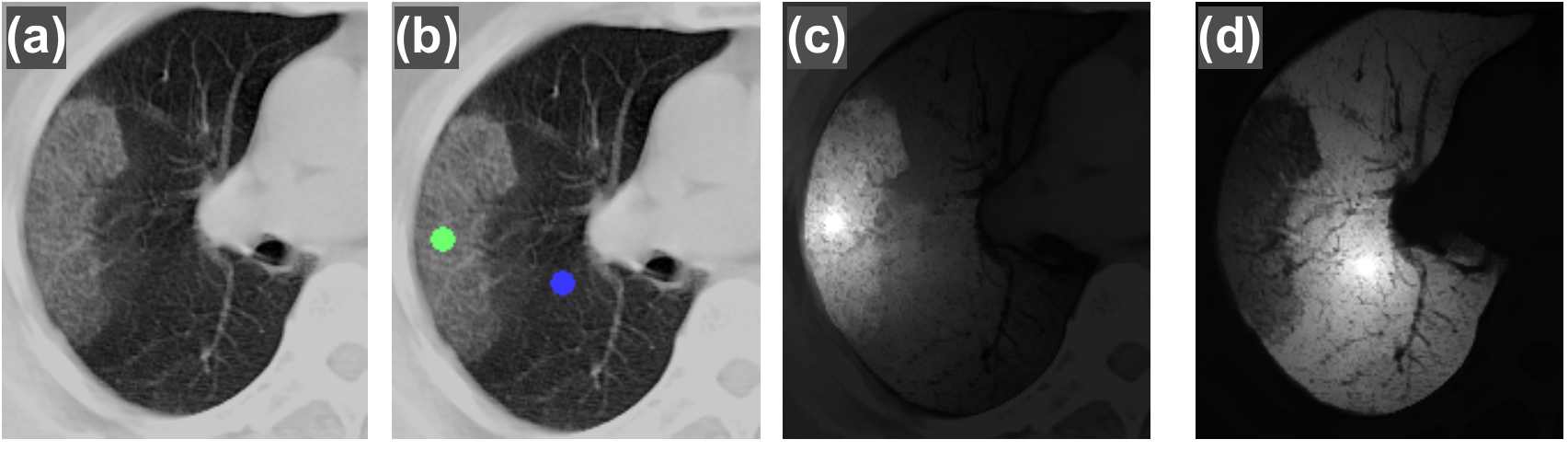}
  \caption{Adaptive online training weights: (a) input, (b) {\color{green}foreground} / {\color{blue} background} scribbles, (c) foreground  and  (d) background weights using $\tau$=$0.2$ in Eq.~\eqref{eq:adaptive_weights}.
  }
  \label{fig:adaptive_weights_intro}
  \vspace{-0.5cm}
\end{figure}
\fi

\subsubsection{Contributions:}
To overcome limitations of existing techniques, we propose adaptive multi-scale online likelihood network (MONet) for AI-assisted interactive segmentation of lung lesions in CT volumes from COVID-19 patients.
Our contributions are three-fold, we propose:
\begin{enumerate}[i.,nosep]
    \item Multi-scale online likelihood network (MONet), consisting of a multi-scale feature extractor, which enables relevant features extraction at different scales for improved accuracy;
    \item Adaptive online loss that uses weights from a scaled negative exponential geodesic distance from user-scribbles, enabling adaptive learning from both initial segmentation and user-provided corrections (\figref{fig:adaptive_weights_intro});
    \item Probability-guided pruning approach where uncertainty from initial segmentation model is used for pruning ambiguous online training data.
\end{enumerate}
MONet enables human-in-the-loop online learning to perform AI-assisted annotations and should not be mistaken for an end-to-end segmentation model.

We perform expert evaluation which shows that adaptively learned MONet outperforms existing state-of-the-art, achieving 5.86\% higher Dice score with 24.67\% less perceived NASA-TLX workload score evaluated.

\section{Method}
\begin{figure}[t!]
  \centering
  \includegraphics[width=1.0\linewidth]{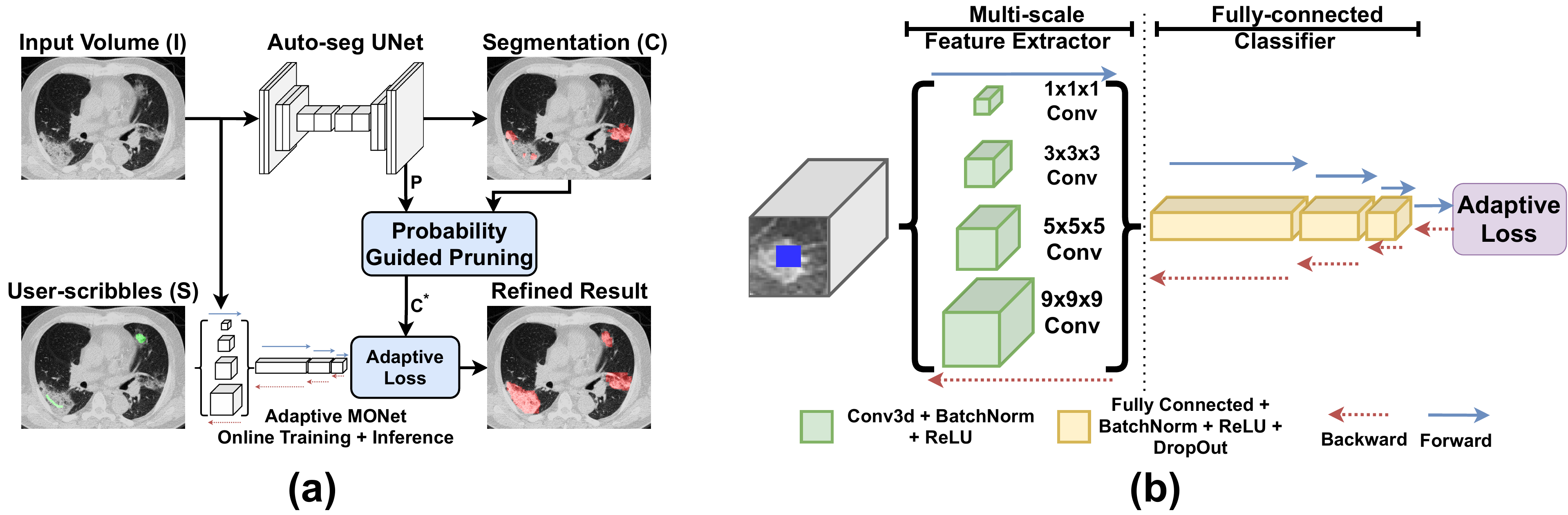}
  \caption{Adaptive learning for interactive segmentation: (a) training and inference of MONet using adaptive loss and probability-guided pruning; (b) architecture of our multi-scale online likelihood network (MONet).}
  \label{fig:MONetFlowchart}
\end{figure}

Given an input image volume, $I$, a pre-trained CNN segmentation model generates an automatic segmentation $C$ with 
associated probabilities $P$. When using data from a new domain, the automated network may fail to properly segment foreground/background objects. To improve this, the user provides scribbles-based interaction indicating corrected class labels for a subset of voxels in the image $I$. Let $\mathcal{S} = \mathcal{S}^f \cup \mathcal{S}^b$ represent these set of scribbles, where $\mathcal{S}^f$ and $\mathcal{S}^b$ denote the foreground and background scribbles, respectively, and $\mathcal{S}^f \cap \mathcal{S}^b = \emptyset$.  \figref{fig:MONetFlowchart}~(a) shows scribbles $\mathcal{S}$, along with the initial segmentation $C$ and probabilities $P$.

\subsection{Multi-scale Online Likelihood Network}
\label{sec:monet_architecture} 
Our proposed multi-scale online likelihood network (MONet), shown in \figref{fig:MONetFlowchart}~(b), uses a multi-scale feature extractor that applies a 3D convolution at various kernel sizes to capture spatial information at different scales. 
The output of each scale is concatenated and fed to a fully-connected classifier, which infers the likelihood for background/foreground classification of the central voxel in the input patch. Each layer in MONet is followed by batch normalization and ReLU activation.

\subsection{Adaptive Loss for Online Learning}
\label{sec:adaptive_online_learning_loss}
The scribbles $\mathcal{S}$ only provide sparse information for online learning. 
However, these corrections are likely also applicable
to
neighboring voxels with similar appearance features,
thereby providing an extended source of training information.
Concurrently, the initial automated segmentation $C$ will often provide reliable results away from the scribbles.
To extend the influence of the scribbles $\mathcal{S}$
while preserving
the quality of the 
initial segmentation $C$, we propose a
spatially-varying
adaptive online loss:
\begin{equation}
  \mathcal{L} = - \sum_i \left[ (1-W_i) \mathcal{L}^{C}_i  + W_i \mathcal{L}^{\mathcal{S}}_i \right],
\label{eq:adaptiveWeightedCrossEntropy}
\end{equation}
where $i$ is a voxel
index,
$\mathcal{L}^C$ and $\mathcal{L}^{\mathcal{S}}$ are individual loss terms for learning from the automated segmentation $C$ and the user-provided correction scribbles $\mathcal{S}$ respectively.
$W$ are spatially-varying interaction-based weights defined using the geodesic distance $D$ between voxel $i$ and the scribbles $\mathcal{S}$:
\begin{align}
 W_i &= \exp\left({\frac{-D(i, S, I)}{\tau}}\right),
 \label{eq:adaptive_weights}
\end{align}
where the temperature term $\tau$ controls the influence of $W$ in $I$.
The
geodesic distance to the scribbles is defined as 
$D(i, \mathcal{S}, I) = {\mathrm{min}}_{j \in \mathcal{S}} \,d(i, j, I)$
\ifnum 1=0
\begin{equation}
    d(i, j, I) = \underset{p \in \mathcal{P}_{i, j}}{\mathrm{min}} \int_{0}^{1} \left \| \nabla I(p(x)) \cdot \mathbf{u}(x)\right \|\,\,dx,
    \label{eq:geo_lambda}
\end{equation}
\else
where 
$d(i, j, I) = {\mathrm{min}}_{p \in \mathcal{P}_{i, j}} \int_{0}^{1} \left \| \nabla I(p(x)) \cdot \mathbf{u}(x)\right \|\,\,dx$
\fi
and
$\mathcal{P}_{i,j}$ is the set of all possible differentiable paths in $I$ between voxels $i$ and $j$.
A feasible path
$p$ is
parameterized by $x \in [0, 1]$.
We denote
$\mathbf{u}(x) = p'(x)/\left\|p'(x)\right\|$ 
the unit vector tangent to the direction of the path $p$.
We further let
$D=\infty$ for $\mathcal{S}=\emptyset$. 

\subsubsection{Dynamic Label-balanced Cross-Entropy Loss:}
\label{sec:class_balanced_loss}
User-scribbles for online interactive segmentation suffer from dynamically changing class imbalance \cite{asad2022econet}.
Moreover, lung lesions in CT volumes usually occupy a small subset of all voxels, introducing additional label imbalance and hence reducing their impact on imbalanced online training. 
To address these challenges, we utilize a label-balanced cross-entropy loss \cite{asad2022econet,ho2019real,kukar1998cost}, with dynamically changing class weights from segmentations and scribbles distribution. 
Given an online model with parameters $\theta$, the foreground likelihood from this model is $p_i = P(s_i=1 | I, \theta)$. 
Then, the segmentations-balanced and scribbles-balanced cross-entropy terms are:
\ifnum 1=0
    {\scriptsize
        \begin{align}
            \label{eq:weightedCrossEntropy}
            \mathcal{L}^{C}_{i} &= \alpha^f y^C_i \log p_i + \alpha^b (1-y^C_i)\log(1-p_i), & 
            \mathcal{L}^{S}_{i} &=\beta^f y^\mathcal{S}_i \log p_i + \beta^b (1-y^\mathcal{S}_i)\log(1-p_i)
        \end{align}
    }
\else
    \begin{equation}
        \label{eq:weightedCrossEntropyC}
        \mathcal{L}^{C}_{i} = \alpha^f y^C_i \log p_i + \alpha^b (1-y^C_i)\log(1-p_i),
    \end{equation}
    \begin{equation}
        \label{eq:weightedCrossEntropyS}
        \mathcal{L}^{S}_{i} = \beta^f y^\mathcal{S}_i \log p_i + \beta^b (1-y^\mathcal{S}_i)\log(1-p_i),
    \end{equation}
\fi
where $\alpha$ and $\beta$ are class weights for labels $C$ and scribbles $\mathcal{S}$ that are defined by labels and scribbles distributions during online interaction as:
%
$\alpha^f = {\left|\mathcal{T}\right|}/{\left| C^f\right|}$,
$\alpha^b = {\left|\mathcal{T}\right|}/{\left| C^b\right|}$,
$\beta^f = {\left|\mathcal{T}\right|}/{\left| \mathcal{S}^f\right|}$,
$\beta^b = {\left|\mathcal{T}\right|}/{\left| \mathcal{S}^b\right|}$ and $\left|\mathcal{T}\right| = \left|C\right| + \left|\mathcal{S}\right|$. $y^C_i$ and $y^\mathcal{S}_i$ represent labels in $C$ and $\mathcal{S}$, respectively. 

The patch-based training approach from \cite{asad2022econet} is used to first extract K$\times$K$\times$K patches from $I$ centered around each voxel in $\mathcal{S}$ and $C$ and train MONet using Eq.~\eqref{eq:adaptiveWeightedCrossEntropy}.
Once learned, efficient online inference from MONet is achieved by applying it to the whole input CT volumes as a fully convolutional network \cite{long2015fully}. 

\subsection{Improving Efficiency with Probability-guided Pruning}
MONet is applied as an online likelihood learning method, where the online training happens with an expert human annotator in the loop, which makes online training efficiency critical. We observe that the automatic segmentation models provide dense labels $C$ which may significantly impact online training and inference performance. $C$ may contain ambiguous predictions for new data, and a number of voxels in $C$ may provide redundant labels. To improve online efficiency while preserving accuracy during training, we prune labels as $C^* = \mathcal{M} \odot C$ where:
\ifnum 1=0
\begin{equation}
  \label{eq:pruning_segmentation}
  \mathcal{M}_i =
  \begin{cases}
    1 & \text{if $P_i \geq \zeta$ and $U_i \geq \eta$}, \\
    0 & \text{otherwise},
  \end{cases}
\end{equation}
\else
$\mathcal{M}_i$ is set to $1$ if $P_i \geq \zeta$ and $U_i \geq \eta$ and $0$ otherwise.
\fi
$\zeta \in [0, 1]$ is the minimum confidence to preserve a label, $U_i \in [0, 1]$ is a uniformly distributed random variable, and $\eta \in [0, 1]$ is the fraction of samples to prune.

\begin{table}[t!]
  \centering
  \caption{State-of-the-art evaluated comparison methods, showing improvement in accuracy (Dice and ASSD) when using different features.  
  Features in \textcolor{blue}{blue} text are proposed in this paper. 
  Key: OL - online learning, PP - post-processing.
  }
  \label{tab:comparison_methods}
  \resizebox{1.0\textwidth}{!}{
    \footnotesize
    \begin{tabular}{L{3.2cm}C{1.5cm}C{1.5cm}C{1.5cm}C{1.5cm}C{1.5cm}C{1.5cm}C{1.5cm}}
    \hline
    \textbf{Method} & \textbf{Technique} & \textbf{Initial} & \textcolor{blue}{\textbf{Multi}} & \textcolor{blue}{\textbf{Adaptive}} & \textcolor{blue}{\textbf{Temp.}} & \textbf{Dice} & \textbf{ASSD} \\
    \textbf{} &  & \textbf{Seg.} & \textcolor{blue}{\textbf{Scale}} & \textcolor{blue}{\textbf{Loss}} & \textcolor{blue}{\textbf{($\tau$)}} & \textbf{(\%)} &  \\ \hline \hline
    \textbf{MONet (proposed)} & \textbf{OL} & \cmark & \cmark & \cmark & \cmark & \textbf{77.77}  &  \textbf{11.82} \\ 
    \textbf{MONet-NoMS} & \textbf{OL} & \cmark & \xmark & \cmark & \cmark & 77.06  & 13.01 \\ 
    \textbf{ECONet\cite{asad2022econet}} & \textbf{OL} & \xmark & \xmark & \xmark & \xmark &  77.02  & 20.19  \\
    \textbf{MIDeepSegTuned\cite{luo2021mideepseg}} & \textbf{PP} & \cmark & \xmark & \xmark & \cmark & 76.00 & 20.16 \\ 
    \textbf{MIDeepSeg\cite{luo2021mideepseg}} & \textbf{PP} & \cmark & \xmark & \xmark & \xmark & 56.85 & 33.25 \\ 
    \textbf{IntGraphCut} & \textbf{PP} & \cmark & \xmark & \xmark & \xmark & 68.58 &  28.64 \\ 
    \hline
  \end{tabular}
  }
\end{table}

\section{Experimental Validation}
\tabref{tab:comparison_methods} outlines the different state-of-the-art interactive segmentation methods and their extended variants that we introduce for fair comparison. We compare our proposed MONet with ECONet \cite{asad2022econet} and MIDeepSeg \cite{luo2021mideepseg}. 
As our proposed Eq.~\eqref{eq:adaptive_weights} is inspired by the exponential geodesic distance from MIDeepSeg \cite{luo2021mideepseg}, we introduce MIDeepSegTuned, which utilizes our proposed addition of a temperature term $\tau$. Moreover, to show the importance of multi-scale features, we include MONet-NoMS which uses features from a single 3D convolution layer. We utilize MONAI Label to implement all online likelihood methods~\cite{diaz2022monai}.
For methods requiring an initial segmentation, we train a 3D UNet~\cite{cciccek20163d} using MONAI \cite{MONAI_Consortium_MONAI_Medical_Open_2020} with features $[32, 32, 64, 128, 256, 32]$. 
Output from each method is regularized using GraphCut optimization. 
We also compare against a baseline interactive GraphCut (IntGraphCut) implementation, that updates UNet output with scribbles based on \cite{boykovjolly2001interactive} and then performs GraphCut optimization. The proposed method is targeted for online training and inference, where quick adaptability with minimal latency is required. Note that incorporating more advanced deep learning methods in this context would result in a considerable decrease in online efficiency, rendering the method impractical for online applications \cite{asad2022econet}.
We utilize a GPU-based implementation of geodesic distance transform \cite{asad2022fastgeodis} in Eq.~\eqref{eq:adaptive_weights}, whereas MIDeepSeg uses a CPU-based implementation. We use NVIDIA Tesla V100 GPU with 32 GB memory for all our experiments. Comparison of accuracy for each method is made using Dice similarity (Dice) and average symmetric surface distance (ASSD) metrics against ground truth annotations \cite{asad2022econet,luo2021mideepseg}. 
Moreover, we compare performance using execution time (Time), including online training and inference time, average full annotation time (FA-Time), and number of voxels with scribbles (S) needed for a given accuracy.

\subsubsection{Data:}
To simulate a scenario where the automatic segmentation model is trained on data from a different source than it is tested on, we utilize two different COVID-19 CT datasets.
The dataset from the COVID-19 CT lesion segmentation challenge \cite{roth2021rapid} is used for training and validation of 3D UNet for automatic segmentation task and patch-based pre-training of MONet/MONet-NoMS/ECONet. This dataset contains binary lung lesions segmentation labels for 199 CT volumes (160 training, 39 validation).
We use UESTC-COVID-19 \cite{wang2020noise}, a dataset from a different source, for the experimental evaluation of interactive segmentation methods (test set).
This dataset contains 120 CT volumes with lesion labels, from which 50 are by expert annotators and 70 are by non-expert annotators. To compare robustness of the proposed method against expert annotators, we only use the 50 expert labelled CT volumes.
\begin{figure}[t!]
\begin{minipage}{0.62\textwidth}
    \centering
      \resizebox{1.0\textwidth}{!}{
        \footnotesize
        \begin{tabular}{L{3.3cm}C{1.9cm}C{1.9cm}C{1.9cm}C{1.4cm}}
        \hline
        \textbf{Method} & \textbf{Dice (\%)} &\textbf{ASSD} &\textbf{Time (s)} &\textbf{Scribbles} \\  \hline \hline
        \textbf{MONet (proposed)} & \textbf{77.77$\pm$6.84}  &  \textbf{11.82$\pm$12.83} & 6.18$\pm$2.42 &  \textbf{20$\pm$24} \\
        \textbf{MONet-NoMS}  & 77.06$\pm$7.27  &  13.01$\pm$15.29 & 7.76$\pm$8.16 &  20$\pm$24 \\ 
        \textbf{ECONet\cite{asad2022econet}} & 77.02$\pm$6.94  & 20.19$\pm$14.71 & 1.46$\pm$1.22 &  2283$\pm$2709 \\ 
        \textbf{MIDeepSegTuned\cite{luo2021mideepseg}} & 76.00$\pm$7.37  &  20.16$\pm$22.57 & 7.97$\pm$2.47 &  23$\pm$17 \\ 
        \textbf{MIDeepSeg\cite{luo2021mideepseg}} & 56.85$\pm$14.25  &  33.25$\pm$25.26 & 6.26$\pm$1.46 &  436$\pm$332 \\ 
        \textbf{IntGraphCut} & 68.58$\pm$9.09  &  28.64$\pm$27.36 & \textbf{0.11$\pm$0.04} &  480$\pm$359 \\ \hline
      \end{tabular}
      }
      \captionof{table}{Quantitative comparison of interactive segmentation methods using synthetic scribbler shows mean and standard deviation of Dice, ASSD, Time and Synthetic Scribbles Voxels.}
      \label{tab:syn_scrib_quant_results}
\end{minipage}
~
\begin{minipage}{0.36\textwidth}
    \centering
    \includegraphics[width=1\linewidth, trim={0.70cm, 1.65cm, 0.6cm, 1.2cm}]{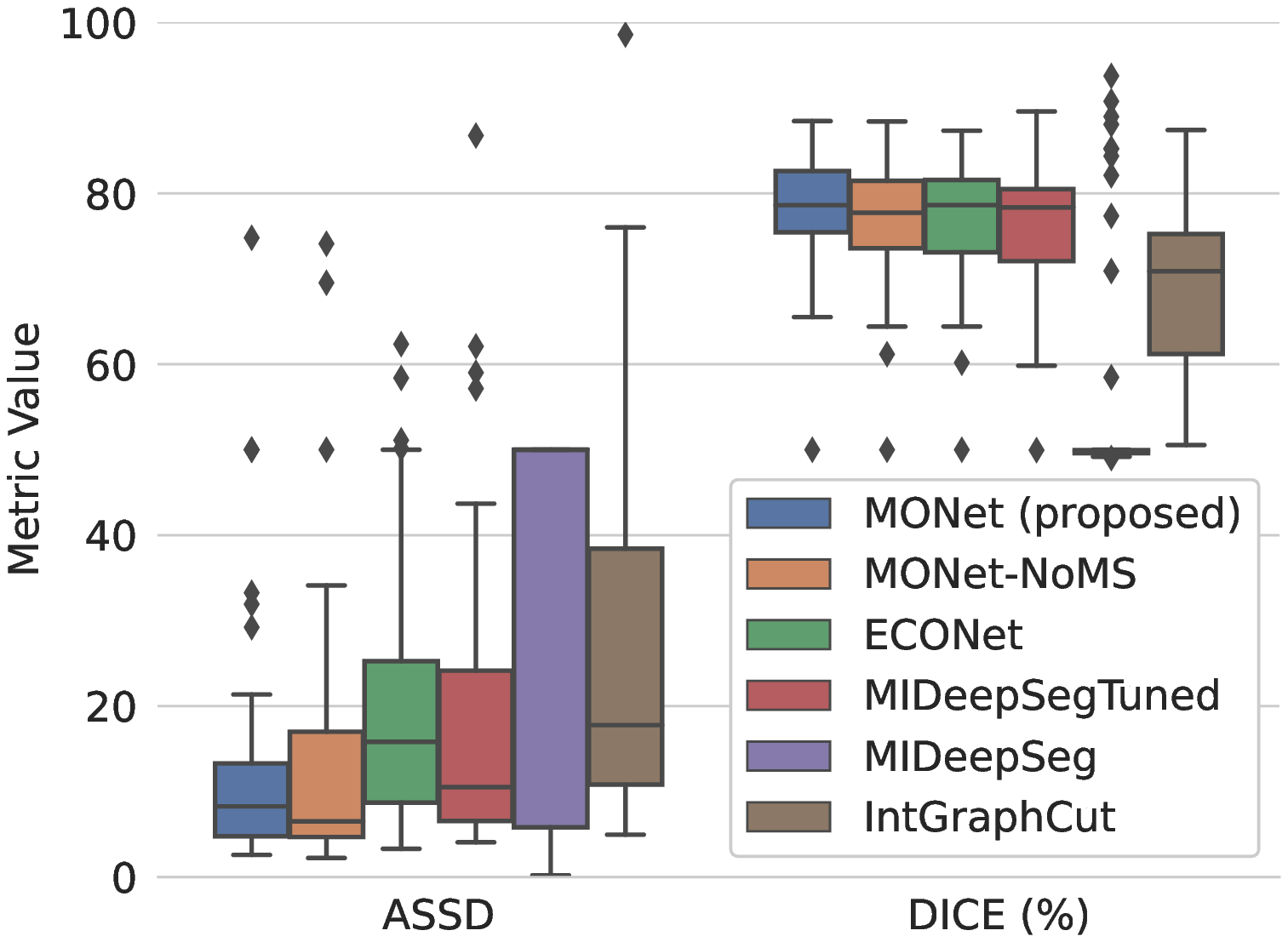}
    \caption{Validation accuracy using synthetic scribbles.} 
    \label{fig:boxplot_accuracy}
\end{minipage}
\end{figure}

\subsubsection{Training Parameters:}
Training of 3D UNet utilized a learning rate (lr) of $1e^{-4}$ for 1000 epochs and MONet/MONet-NoMS offline pre-training used 50 epochs, and $\text{lr}=1e^{-3}$ dropped by 0.1 at the 35th and 45th epoch. Online training for MONet, MONet-NoMS and ECONet \cite{asad2022econet} used 200 epochs with $\text{lr}=1e^{-2}$ set using cosine annealing scheduler \cite{loshchilov2016sgdr}. Dropout of 0.3 was used for all fully-connected layers in online models. Each layer size in ECONet and MONet-NoMS was selected by repeating line search experiments from \cite{asad2022econet}: (i) input patch/3D convolution kernel size of $K=9$, (ii) 128 input 3D convolution filters and (iii) fully-connected sizes of 32$\times$16$\times$2. For MONet, we utilize four input 3D convolution with multi-scale kernel sizes $K = [1, 3, 5, 9]$ with each containing 32 filters (i.e., a total of 128 filters, same as (ii)). We utilize the same fully-connected sizes as in (iii) above. 
Parameters $\zeta=0.8$ and $\eta=0.98$ are selected empirically. We utilize $\tau=0.3$ for MONet, MIDeepSegTuned and MONet-NoMS.
We use GraphCut regularization, where $\lambda=2.5$ and $\sigma=0.15$ \cite{boykovjolly2001interactive}. 
Search experiments used for selecting $\tau$, $\lambda$, $\sigma$ are shown in Fig.~1 and 2 in supplementary material.


\subsection{Quantitative Comparison using Synthetic Scribbler}
\label{sec:quant_syntheticscrib}
We employ the synthetic scribbler method from \cite{asad2022econet,wang2018deepigeos} where mis-segmented regions in the inferred segmentations are identified by comparison to the ground truth segmentations. 
\tabref{tab:syn_scrib_quant_results} and \figref{fig:boxplot_accuracy} present quantitative comparison of methods using synthetic scribbler. They show that MONet outperforms all existing state-of-the-art in terms of accuracy with the least number of synthetic scribbled voxels. In particular, MONet outperforms both MIDeepSeg \cite{luo2021mideepseg} and MIDeepSegTuned, where adaptive online learning enables it to quickly adapt and refine segmentations.    
In terms of efficiency, online training and inference of the proposed MONet takes around 6.18 seconds combined, which is 22.4\% faster as compared to 7.97 seconds for MIDeepSeg. However, it is slower than ECONet and ISeg. MIDeepSeg performs the worst as it is unable to adapt to large variations and ambiguity within lung lesions from COVID-19 patients, whereas by utilizing our proposed  Eq.~\eqref{eq:adaptive_weights} in MIDeepSegTuned, we improve its accuracy. When comparing to online learning methods, MONet outperforms MONet-NoMS, where the accuracy is improved due to MONet's ability to extract multi-scale features. Existing state-of-the-art online method ECONet \cite{asad2022econet} requires significantly more scribbled voxels as it only relies on user-scribbles for online learning.

\begin{figure}[t!]
\begin{minipage}{0.4394\textwidth}
    \centering
      \resizebox{1.0\textwidth}{!}{
        \footnotesize
        \begin{tabular}{L{2.00cm}|C{1.8cm}C{2.0cm}}
            \hline
            & \textbf{MONet (proposed)} & \textbf{MIDeepSeg-Tuned\cite{luo2021mideepseg}} \\ \hline \hline
            \textbf{Finished} & \textbf{100\%} & 33.33 \% \\
            \textbf{NASA-TLX} & \textbf{52.33} & 77.00 \\
            \textbf{Dice (\%)} & \textbf{88.53$\pm$2.27} & 82.67$\pm$12.36 \\
            \textbf{ASSD} & \textbf{2.91$\pm$1.58} & 11.30$\pm$20.09 \\
            \textbf{FA-Time (s)} & \textbf{507.11} & 567.43 \\ \hline
      \end{tabular}
      }
       \captionof{table}{Workload validation by expert user, shows Dice (\%), ASSD, full annotation time, FA-Time (s), overall NASA-TLX perceived workload score and the \% of data successfully annotated by expert.
      }
      \label{tab:nasa_tlx_study_compare}
\end{minipage}
~
\begin{minipage}{0.5594\textwidth}
    \centering
      \resizebox{0.9\textwidth}{!}{
        \footnotesize
        \begin{tabular}{C{3.1cm}|C{1.8cm}C{2.0cm}}
            \hline
            \textbf{NASA-TLX weighted scores}  & \textbf{MONet (proposed)} & \textbf{MIDeepSeg-Tuned\cite{luo2021mideepseg}} \\ \hline \hline
            \textbf{Effort}          & \textbf{14.67} & 21.33   \\
            \textbf{Frustration}     & \textbf{7.67} & 16.67  \\
            \textbf{Mental Demand}   & \textbf{13.33} & 15.00  \\
            \textbf{Performance}     & \textbf{10.00} & 17.00   \\
            \textbf{Physical Demand} & \textbf{4.67} & 7.00     \\
            \textbf{Temporal Demand} & 2.00 & \textbf{0.00}     \\ \hline
            \textbf{Total workload}  & \textbf{52.33} & 77.00    \\ \hline
        \end{tabular}
      }
      \captionof{table}{NASA-TLX perceived workload by expert user, shows total workload and individual sub-scale scores. The method with low score requires less effort, frustration, mental, temporal and physical demands with high perceived performance. 
      }
      \label{tab:nasa_tlx_raw}
    \end{minipage}
\end{figure}

\subsection{Performance and Workload Validation by Expert User}
\label{sec:userstudy}
This experiment aims to compare the 
performance and
\emph{perceived} subjective workload of the proposed MONet with the best performing comparison method MIDeepSegTuned based on \cite{luo2021mideepseg}. We asked an expert, with 2 years of experience in lung lesion CT from 
\ifdefined\ANONYMIZED
X,
\else
Radiology Department, Oxford University Hospitals NHS Foundation Trust, 
\fi
to utilize each method for labelling the following pathologies as lung lesions in 10 CT volumes from UESTC-COVID-19 expert set \cite{wang2020noise}: ground glass opacity, consolidation, crazy-paving, linear opacities. One CT volume is used by the expert to practice usage of our tool.
The remaining 9 CT volumes were presented in a random order, where the perceived workload was evaluated by the expert at half way (after 5 segmentations) and at the end. We use the National Aeronautics and Space Administration Task Load Index (NASA-TLX) \cite{hart2006nasa}
as per previous interactive segmentation studies
\cite{mcgrath2020manual,ramkumar2017using,williams2021interactive}. The NASA-TLX asks the expert to rate the task based on six factors, being performance, frustration, effort, mental, physical and temporal demand. The weighted NASA-TLX score is then recorded as the expert answers 15 pair-wise questions rating factors based on importance. In addition, we also recorded accuracy metrics (Dice and ASSD) against ground truth labels in \cite{wang2020noise}, time taken to complete annotation and whether the expert was able to successfully complete their task within 10 minutes allocated for each volume. 

\begin{figure}[t!]
  \centering
  \includegraphics[width=1.0\linewidth]{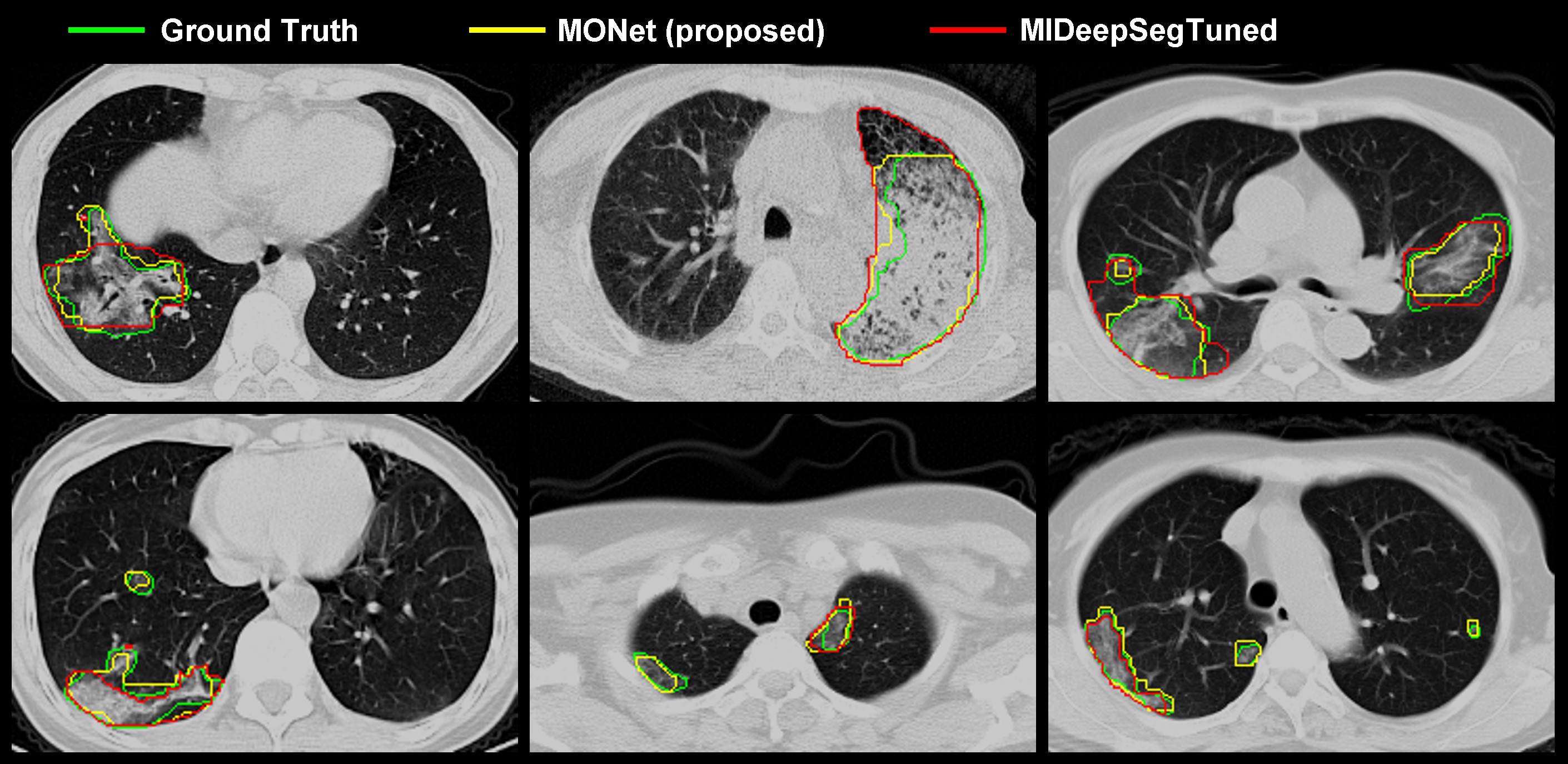}
  \caption{Visual comparison of interactive segmentation results from \secref{sec:userstudy}. Segmentations are shown with contours on axial plane slices from different cases. 
  }
  \label{fig:user_scrib_qualit_results_analysis}
\end{figure}

\tabref{tab:nasa_tlx_study_compare} presents an overview for this experiment, where using the proposed MONet, the expert was able to complete 100\% of the labelling task, whereas using MIDeepSegTuned they only completed 33.33\%
within the allocated time.
In addition, MONet achieves better accuracy with lower time for complete annotation and less overall perceived workload with NASA-TLX of 52.33\% as compared to 77.00\% for MIDeepSegTuned. 
\tabref{tab:nasa_tlx_raw} shows the individual scores that contribute to overall perceived workload. It shows that using the proposed MONet, the expert perceived reduced workload in all sub-scale scores except temporal demand. We believe this is due to the additional online training/inference overhead for MONet application.
\figref{fig:user_scrib_qualit_results_analysis} visually compares these results where MONet results in more accurate segmentation as compared to MIDeepSegTuned. We also note that MONet's ability to apply learned knowledge on the whole volume enables it to also infer small isolated lesions, which MIDeepSegTuned fails to identify.

\section{Conclusion}
We proposed a multi-scale online likelihood network (MONet) for scribbles-based AI-assisted interactive segmentation of lung lesions in CT volumes from COVID-19 patients. MONet consisted of a multi-scale feature extractor that enabled extraction of relevant features at different scales for improved accuracy. We proposed an adaptive online loss that utilized adaptive weights based on user-provided scribbles that enabled adaptive learning from both an initial automated segmentation and user-provided label corrections. Additionally, we proposed a dynamic label-balanced cross-entropy loss that addressed dynamic class imbalance, an inherent challenge for online interactive segmentation methods. Experimental validation showed that the proposed MONet outperformed the existing state-of-the-art on the task of annotating lung lesions in COVID-19 patients. Validation by an expert showed that the proposed MONet achieved on average 5.86\% higher Dice while achieving 24.67\% less perceived NASA-TLX workload score than the MIDeepSegTuned method \cite{luo2021mideepseg}. 

\section{Acknowledgments}
This project has received funding from the European Union's Horizon 2020 research and innovation programme under grant agreement No 101016131 (icovid project). This work was also supported by core and project funding from the Wellcome/EPSRC [WT203148/Z/16/Z; NS/A000049/1; WT101957; NS/A000027/1]. This project utilized scribbles-based interactive segmentation tools from opensource project MONAI Label \footnote{\url{https://github.com/Project-MONAI/MONAILabel}} \cite{diaz2022monai}.

%
\printbibliography

\end{document}



  
%
\title{Supplementary Material: Adaptive Multi-scale Online Likelihood Network for AI-assisted Interactive Segmentation}
%
\titlerunning{Supplementary Material: Adaptive Multi-scale Online Likelihood Network}
\ifdefined\ANONYMIZED
\author{Anonymous MICCAI Submission, Paper 2125}
\else
\author{Muhammad Asad\inst{1}
\and
Helena Williams \inst{2}
\and
Indrajeet Mandal\inst{3} \and
Sarim Ather \inst{3} \and
Jan Deprest \inst{2} \and
Jan D'hooge \inst{4} \and
Tom Vercauteren \inst{1}
}
\authorrunning{M Asad et al.}

%
\institute{School of Biomedical Engineering \& Imaging Sciences, King’s College London, UK \and Department of Development \& Regeneration, KU Leuven, Belgium \and Radiology Department, Oxford University Hospitals NHS Foundation Trust, UK \and
Department of Cardiovascular Sciences, KU Leuven, Belgium
}
\fi

\maketitle              
\begin{figure}[htb!]
  \centering
  \includegraphics[width=0.75\linewidth,trim={0cm, 1cm, 0cm, 1cm}]{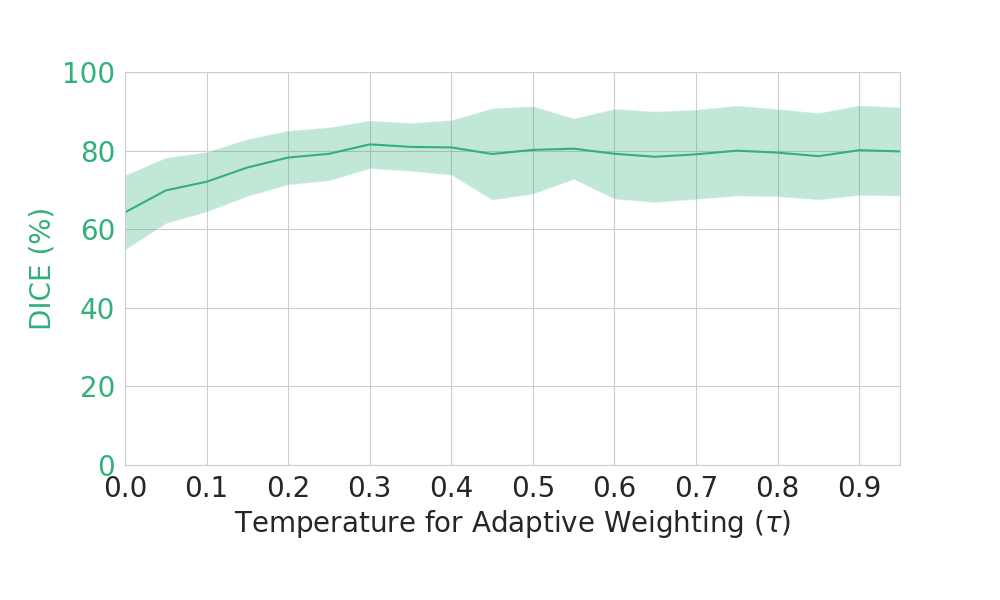}
  \caption{
  Line search for finding optimal $\tau$ in Eq.~(2) in the paper. Shows mean and standard deviation of Dice (\%) accuracy for a given $\tau \in [0,1]$. The selected optimal value $\tau=0.3$ results in the best Dice with the least standard deviation.
  }
  \label{fig:syneval_accuracy1}
\end{figure}

\begin{figure}[htb!]
  \centering
  \includegraphics[width=0.8\linewidth,trim={0cm, 2cm, 0cm, 2cm}]{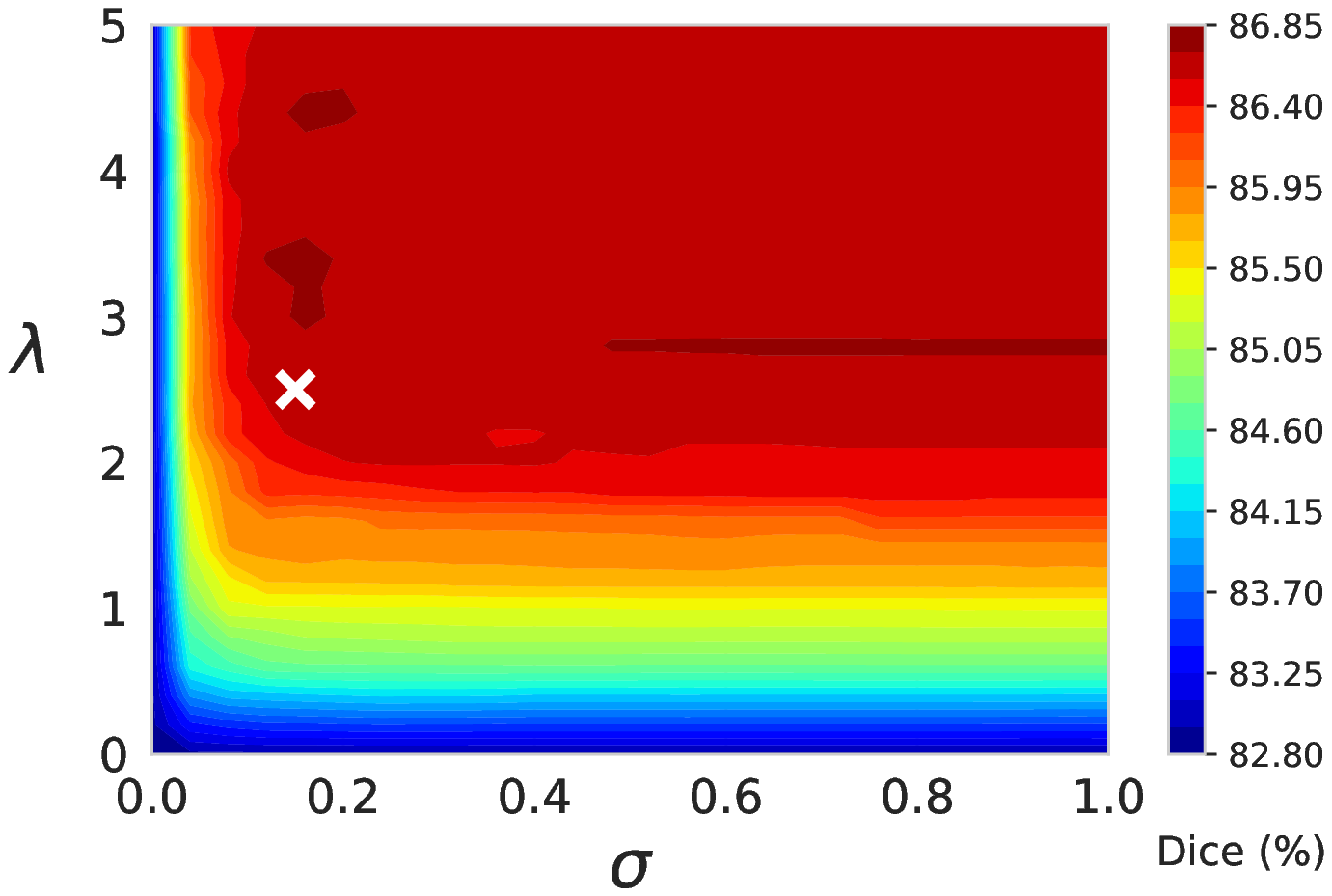}
  \caption{
  Grid search for finding the optimal parameters for GraphCut regularization, where $\lambda$ and $\sigma$ are shown along with corresponding Dice (\%) accuracy. $\times$ shows the optimal values ($\lambda=2.5$, $\sigma=0.15$) used in our paper experiments.
  }
  \label{fig:syneval_accuracy2}
\end{figure}

\begin{figure}[t!]
  \centering
  \includegraphics[width=1.0\linewidth]{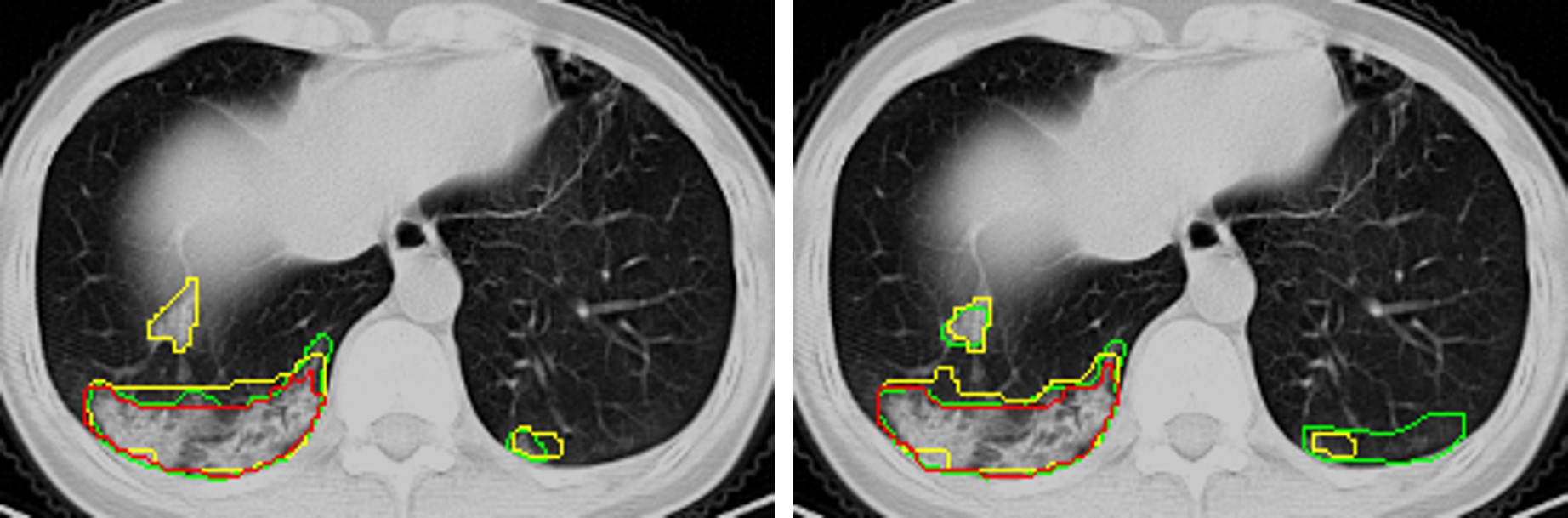}
  \caption{
  Visual comparison of two consecutive slices showing results using scribbles from an expert annotator with MONet and MIDeepSegTuned. The advantage of using proposed MONet is visible where it even addresses discrepancies within ground truth labels by applying learned knowledge to infer possible lesion and non-lesion segmentations. 
  }
  \label{fig:user_scrib_qualit_results_analysis}
\end{figure}

%
%
%